\documentclass[cits]{PoS}

\usepackage{epsfig}
\usepackage{amsmath}
\usepackage{amssymb}
\usepackage{xspace}

\newcommand{\ca}{\ensuremath{C_{\!A}}\xspace}
\newcommand{\cf}{\ensuremath{C_{\!F}}\xspace}
\newcommand{\Nc}{\ensuremath{N_{\!C}}\xspace}
\newcommand{\gs}{\ensuremath{g}\xspace}

\title{
  \vspace{-2cm}
  \begin{flushright}
  \it \normalsize CERN-PH-TH/2010-007
  \end{flushright}
 \vspace{16mm}
All-Order Corrections and Multi-Jet Rates}

\ShortTitle{All-Order Corrections and Multi-Jet Rates}

\author{\speaker{Jeppe R.~Andersen}%
        \\Theory Division, Physics Department, CERN, CH-1211 Geneva 23, Switzerland\\
        E-mail: \email{jeppe.andersen@cern.ch}}

\author{Jennifer M.~Smillie\\
        Department of Physics and Astronomy, UCL, Gower Street, London, WC1E 6BT, UK\\
        E-mail: \email{smillie@hep.ucl.ac.uk}}

\abstract{We discuss results from a recently proposed all-order description
  of hard, radiative corrections to certain multi-jet processes at hadron
  colliders. The description is based on obtaining an all-order estimate of the
  $t$-channel singularities of scattering amplitudes. As a simple example, we
  illustrate the similarities between $qQ$ and $qg$-scattering. In
  particular, we discuss how at tree-level, all non-suppressed
  helicity-amplitudes for these processes consist of a pure $t$-channel
  pole. This structure is used in the construction of all-order approximations.}

\FullConference{RADCOR 2009 - 9th International Symposium on Radiative Corrections (Applications of Quantum Field Theory to Phenomenology) \\
		 October 25-30 2009\\
		 Ascona, Switzerland}

\begin{document}

\section{Introduction}
\begin{figure}[!hbtp]
  \centering
  \epsfig{width=0.45\textwidth,file=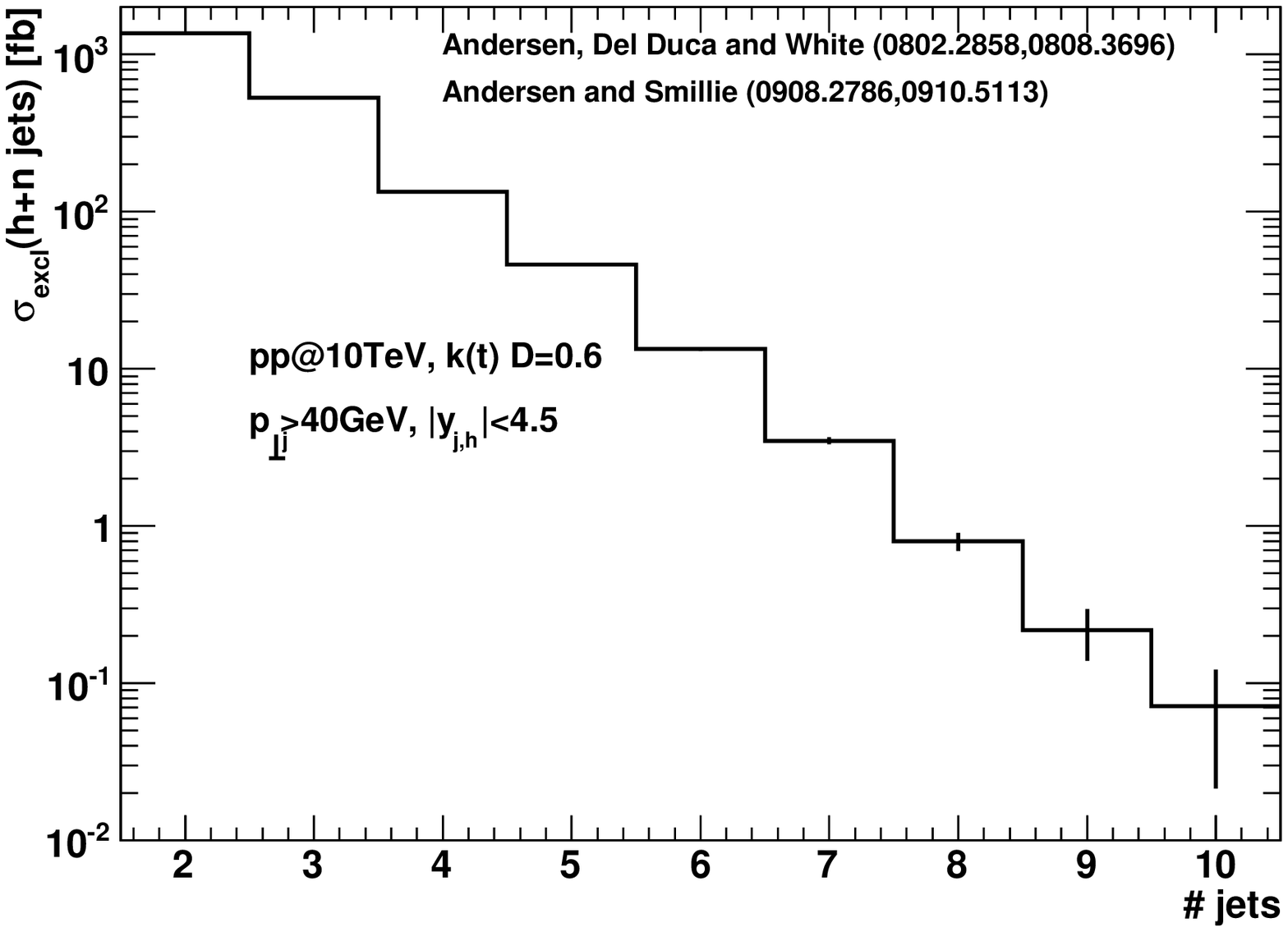}
  \epsfig{width=0.45\textwidth,file=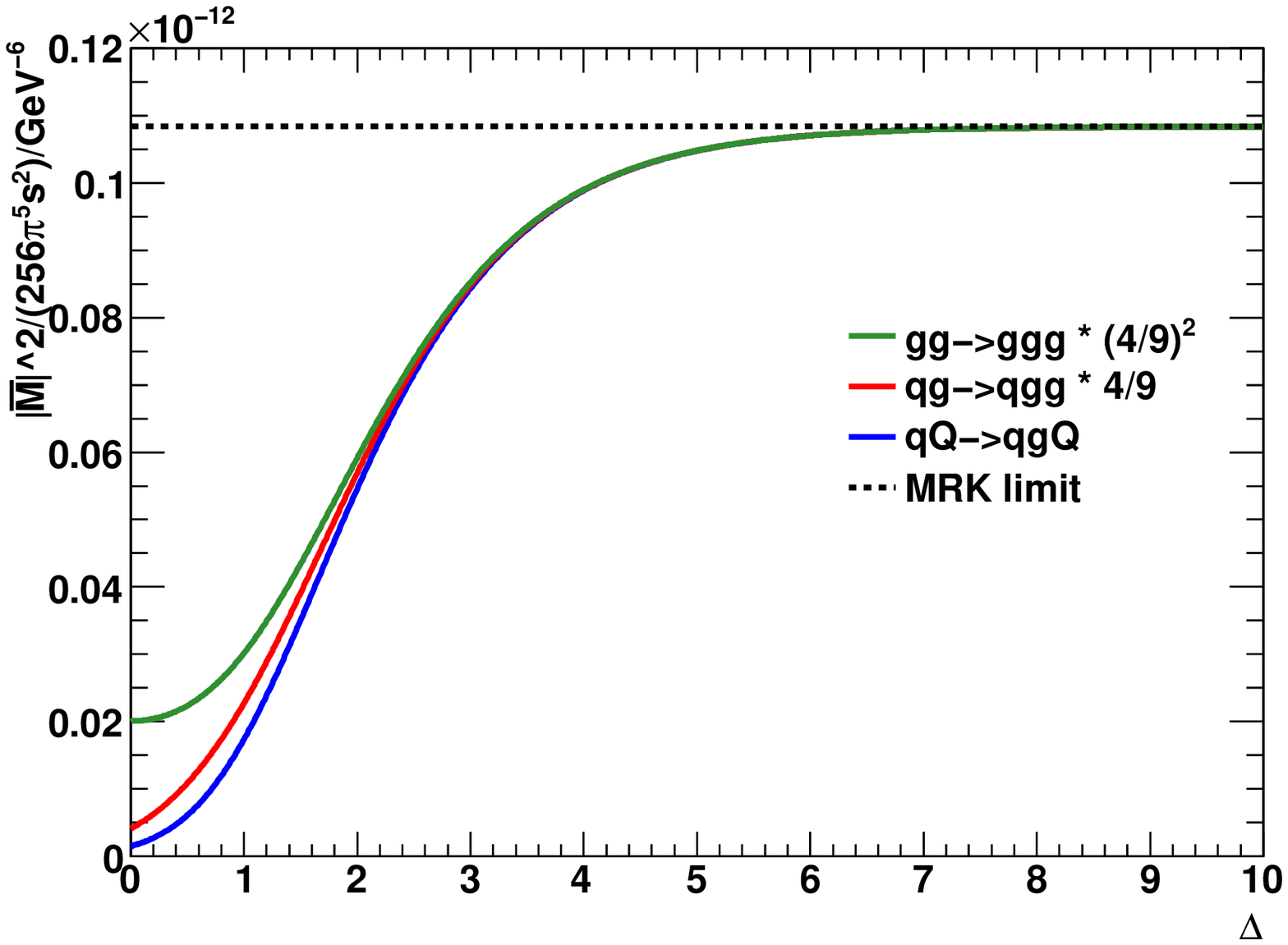}

  (a) \hspace{6cm} (b)
  \caption{(a) The exclusive $n$-jet rates in $pp(10\mathrm{TeV})\to hjj$, as
    obtained in the framework of
    Ref.\cite{Andersen:2008ue,Andersen:2008gc,Andersen:2009nu,Andersen:2009he}. Monte
    Carlo uncertainty only. The uncertainty induced by the choice of
    renormalisation and factorisation scale could be evaluated directly like
    in Ref.\cite{Andersen:2008gc}. (b) $|\mathcal{M}|^2/(256 \pi^5\hat s^2)$ at tree-level
    for $qQ\to qgQ$, $qg\to qgg$ and $gg\to ggg$ with the partons at
    $-\Delta,0,\Delta$ in rapidity, compared with the approximation obtained
    from the extreme MRK limit.}
  \label{fig:HNjets}
\end{figure}

%% Introduction: The problem of predicting the effect of all-order, hard
%%               radiation. Necessary at the LHC. Shower+Tree,... alternative
The description of the hard perturbative corrections to any given scattering
process at colliders is receiving increasing attention, necessitated by both
the ever increasing number of jets entering the search channels for physics
beyond the Standard Model, and the natural increasing rate at which hard
pertubative corrections appear, as the energy of the collider is
increased. In contrast to the situation at earlier colliders, hard
pertubative corrections, giving rise to extra identifiable hard jets, will
play a very significant r\^ole at the LHC. This is because the increase in
energy allows the exploration of more of the hard multi-jet phase space,
which was previously inaccessible. As an example, we illustrate in
Fig.~\ref{fig:HNjets} the \emph{exclusive} jet rates obtained in an all-order
calculation of the radiative corrections to the process $pp\to hjj$ using the
following minimal cuts: $k_t$-algorithm\cite{Cacciari:2005hq} with D=0.6,
$p_{t,\mathrm{jet}}>40$~GeV, $|y_{j,h}|<4.5$. We see that the emission of
each hard jet in this inclusive sample is suppressed by roughly a factor of 3-4. 
This ratio between the rate for a Higgs boson in association with two or
three jets is consistent with estimates based on just (the inclusive)
tree-level calculations of these processes. With such high rates for
additional hard radiation, it is important to take these radiative
corrections into account not just when estimating total rates, but especially
when performing a detailed analysis of the final state, which will be needed
in order to extract e.g.~the $CP$-properties of the Higgs boson. 

The focus is not just on radiative corrections sufficiently hard to form
additional jets, but, especially when not just the total rate but the jet
topology is discussed, also the (semi-) hard radiation just below the jet
scale, which nevertheless changes the orientation of the hard jets. A
next-to-leading order calculation (the current limit whenever jets are
involved) would include the first contribution to these effects. However,
when studying the details of the final state for e.g.~$W+\ge n$-jets, one
finds a strong correlation between the jet emission and the rapidity span of
the event. See e.g.~Ref.\cite{Berger:2009dq} fig.~9 for $n=2$, where it is
evident that when analysing the region of increasing rapidity span, the 3-jet
rate is much more significant than in completely inclusive analyses. This
effect is not special to $W$+jets, but will be seen for all processes with
more than two jet, where these can exchange a
gluon\cite{Andersen:2001ja,Andersen:2003gs,Andersen:2008gc}. An estimate of
even higher order effects is mandatory here to obtain a reliable description
of the configuration of final state jets. Such an estimate can be obtained
using parton shower Monte Carlos. However, this description concentrates on
soft and collinear emissions, and is known to severely underestimate the
effects of hard emissions. In the current contribution, we will discuss a new
approach to all order corrections in processes with at least two hard
jets\cite{Andersen:2008ue,Andersen:2008gc,Andersen:2009nu,Andersen:2009he},
inspired by the behaviour of such scattering amplitudes at large invariant
mass between all produced particles, the so-called Multi-Regge-Kinematic
limit\cite{Lipatov:1974qm}. The formalism takes
into account not just real emissions, but also the (logarithmically) leading
virtual corrections in this limit, thus allowing an IR finite, all-order
summation of radiative corrections.

\section{The High Energy Limit of Scattering Amplitudes from Current-Current Scattering}
\label{sec:high-energy-limit}
%% The HE Limit: MRK limit of amplitudes are known, but do not give a
%%               satisfactory description when applied in all of phase
%%               space. Build better formalism.
%%               Reconstruct t-channel singularities; numerators and
%%               denominators. 
%%               Similarities of channels: qQ easy. qg surprisingly similar
The behaviour of $n$-jet production amplitudes is known in the very exclusive
limit of large invariant mass between each and every jet of comparable and
fixed transverse momentum (not increasing with the centre of mass energy
$\hat s$). In this limit, the dominant contribution is through partonic
channels, which allow for a colour octet exchange between each neighbouring
(in rapidity) parton pair. In the limit, such amplitudes contribute a factor
$4\gs^2\ca/{k_{i\perp}^2}$ to the square of the amplitude. This leads to the
following asymptotic form of the square of the tree-level scattering matrix
element for the processes $gg\to g\cdots g$:
\begin{align}
  \label{eq:Msqsumavg}
  \left|\overline{\mathcal{M}}_{gg\to g\cdots g}^{MRK}\right|^2 = \frac {4\
    \hat s^2}{\Nc^2-1}
  \ \frac{g^2\ \ca}{|p_{1\perp}|^2} \left( \prod_{i=2}^{n-1}\frac{4\ g^2
      \ca}{|p_{i\perp}|^2}\right) \frac{g^2\ \ca}{|p_{n\perp}|^2}.
\end{align}
The momenta $p_1,\ldots,p_n$ are here ordered according to decreasing
rapidity. The same limit of $n$-jet production in $qg$-scattering
($qQ$-scattering) is found by replacing one (two) $\ca\to\cf$. All rapidity
dependence has disappeared in Eq.~\eqref{eq:Msqsumavg} (since the limit of
infinite rapidity separation between all particles has been applied).

In Fig.~\ref{fig:HNjets}(b) we compare $|M|^2/\hat s^2$ for $qQ\to
qgQ$, $qg\to qgg$ and $gg\to ggg$ (the last two rescaled with appropriate
powers of $\ca/\cf$) for a ``Mercedes star'' azimuthal configuration of three
40~GeV jets, with the rapidities of the three jets chosen as
$-\Delta,0,\Delta$ respectively (see Ref.\cite{Andersen:2009nu} for further details).
It is immediately clear that 1) the square of the matrix element does tend to
the right MRK limit, and 2) applying the approximate form of
Eq.~\eqref{eq:Msqsumavg} in all of phase space will lead to a very poor
approximation of the scattering cross section.%\footnote{However, the use of
%  this approximation forms the basis of the BFKL equation}.
% Unfortunately, as can be seen above, this limit significantly overestimates the matrix
% elements in all relevant phase space for any collider.
% However, as can also be seen in Figure \ref{fig:MRK}, after multiplication by the
% appropriate colour factor, the different channels exhibit
% `universal' behaviour at far lower rapidity spans.  This suggests that all channels can be
% successfully modelled in the same way, as we will show.
% Throughout, we will label particle momenta in order of decreasing rapidity as $p_a p_b\to
% p_1\ldots p_n$.  
The inability of the MRK approximation to describe the form of the full
matrix element at rapidities and energies of interest to colliders is caused
by some of the approximations to the kinematic invariants necessary to cast
everything in terms of transverse components only. 

In the following, we will describe recent
progress\cite{Andersen:2009nu,Andersen:2009he} in obtaining a simple
approximation to the all-order $n$-jet cross section (also in association
with a $W,Z$ or $h$-boson), which reproduces the full result in the MRK
limit, but avoids many of the approximations which spoils the straightforward
applicability of Eq.~\eqref{eq:Msqsumavg}. The basic idea is to obtain a
gauge invariant approximation to the $t$-channel singularities of a given
scattering amplitude. We will see how that can be achieved by analysing directly
the structure of amplitudes (rather than the square of these).

Let us therefore begin by studying the simple process of $q^-Q^-\to q^-Q^-$,
where $q$ and $Q$ represent different quark flavours and the negative signs
represent helicity (obviously all helicity configurations lead to equally
simple expressions). The only diagram is $t$-channel gluon exchange giving
the following (colour and coupling stripped) matrix element:
\begin{align}
  \label{MqQ}
  \mathcal{M}_{q^- Q^- \to q^- Q^-}=\left(\bar u_1 \gamma^\mu u_a\right)\ \frac{g_{\mu \nu}}{t}\ \left(\bar u_2 \gamma^\nu u_b\right).
\end{align}
This naturally already has an attractive form ``factorised'' into two
currents, one which depends on the momenta of $q$ only ($a$ and 1), and one
which depends on those of $Q$ only ($b$ and 2). This factorised form is
convenient also for analysing $qg$-scattering, and generalising to $n$-parton
processes (see Ref.\cite{Andersen:2009nu} for further details). A
re-arrangement of this by use of Fierz-identities would lead to
spinor-products depending on the momenta of both q and Q; kinematic
approximations of such spinor products would then be necessary in order to
recast the (then approximated) matrix element on a factorised form (such a
procedure was followed in e.g.~Ref\cite{DelDuca:1999ha}), and would lead to
gauge-dependence when applied to gluon scattering in the sub-MRK region.

Consider now the process $q^- g^- \to q^- g^-$, which has $s$, $t$ and
$u$-channel contributions. However, the full, gauge-invariant scattering
amplitude can be written as~\cite{Andersen:2009he}
\begin{align}
  \label{Mqg}
  \mathcal{M}_{q^- g^- \to q^- g^-} = e^{i\phi_2} \left( t_{ae}^2 t_{e1}^b \sqrt{\frac{p_b^-}{p_2^-}}
      - t^b_{ae}t^2_{e1}\sqrt{\frac{p_2^-}{p_b^-}} \right)\mathcal{M}_{q^- Q^- \to q^- Q^-},
\end{align}
where $p_2\ (p_b)$ is the momentum of the outgoing (incoming) gluon, which we
wlog.~have taken along the negative light-cone direction, $t_{ij}^a$ are
colour matrices, and $p_i^-=p_{i_E}-p_{i_z}$ and $e^{i\phi_2}$ is the phase
of $(p_{2_x}+ip_{2_y})$. Again we find a fully factorised structure with just
a $t$-channel pole: The difference compared with the amplitude for pure quark
scattering is a factor depending on the gluon momenta only. At amplitude
squared level, the result is the same as for $qQ$-scattering, but with a
factor of \cf replaced with
\begin{align}
  \label{CAM}
  \frac12 \left(\ca-\frac1\ca \right) \left(\frac{p_b^-}{p_2^-}+\frac{p_2^-}{p_b^-}
  \right) + \frac1\ca.
\end{align}
This momenta-dependent colour factor gives a measure of how the strength of
the gluon current increases with increasing acceleration of the scattering
gluon, and so we call Eq.~(\ref{CAM}) the \emph{Colour Acceleration
  Multiplier} (CAM). The bigger the difference between the light-cone momenta
of the incoming and outgoing gluon, the stronger the enhancement over pure
quark scattering. One notes that the expression in Eq.~\eqref{CAM} tends to
\ca in the limit $p_2^-\to p_b^-$ (i.e.~the MRK limit), and the $qg$
scattering amplitudes asymptotes to that of a rescaled
$qQ$-scattering. Applications of Fierz identities would again have spoilt
this easily identifiable factorisation between a quark current, and a
(scaled) current depending on gluon momenta only. This is the form one finds
for all of the scattering amplitudes, where the helicity of the gluon is
conserved. All these amplitudes have just a $t$-channel pole. And all other
helicity configurations are systematically suppressed by $\hat s$ in the MRK
limit.

However, even for the amplitudes where the gluon helicity is flipped can the
$t$-channel pole be expressed as quark currents contracted with a quantity
depending on the momenta of the gluon only. Furthermore, the $u$ and
$s$-channel pole is isolated to the case where the helicity of the incoming
quark and gluon is the same, but the helicity of the outgoing gluon is
flipped. Contributions unrelated to the $t$-channel pole are clearly
suppressed as $\hat s/\hat t\to\infty$, and will be ignored in the following.

\begin{figure}[!hbtp]
  \centering
  \epsfig{width=0.45\textwidth,file=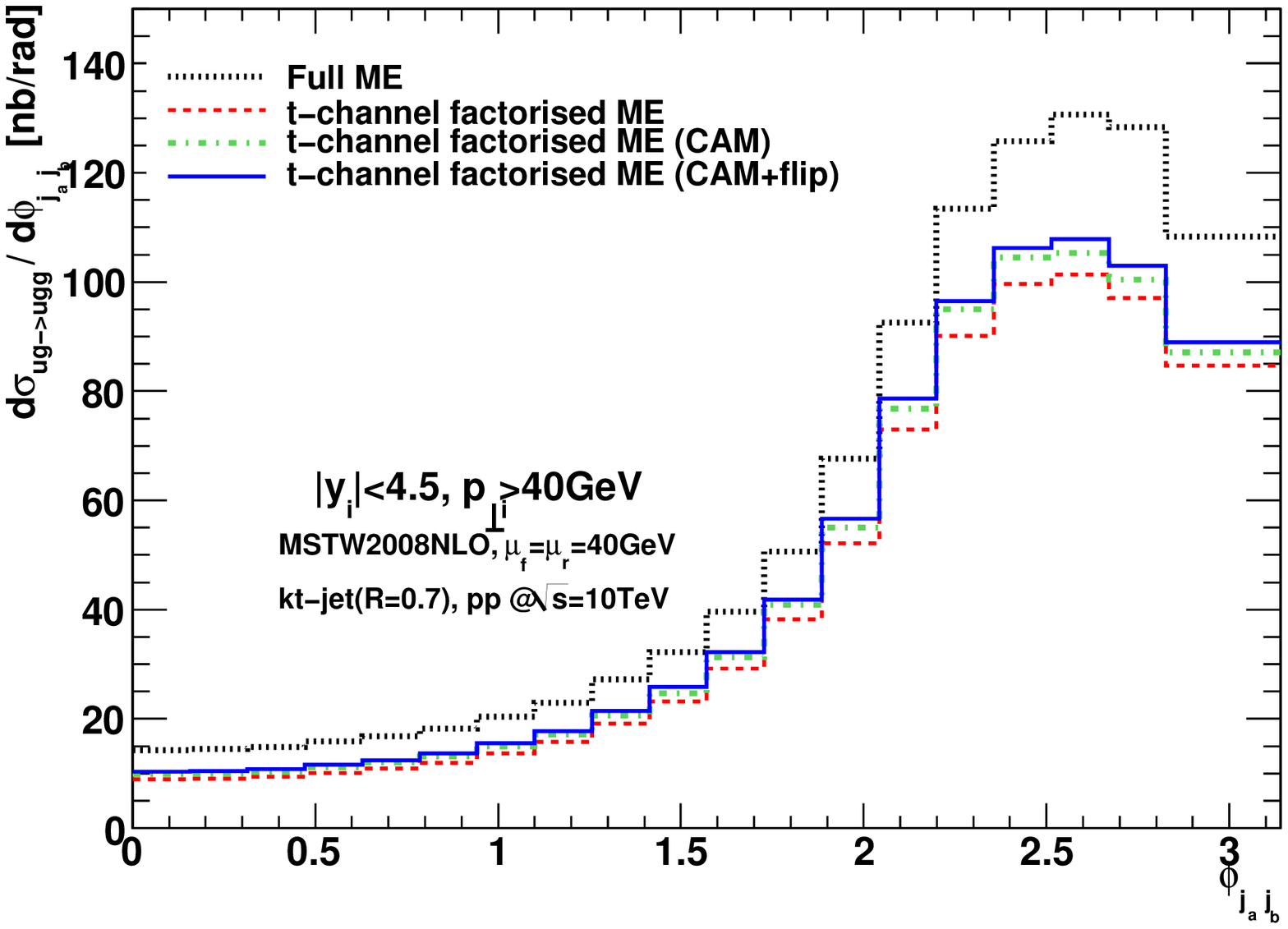}
  \epsfig{width=0.45\textwidth,file=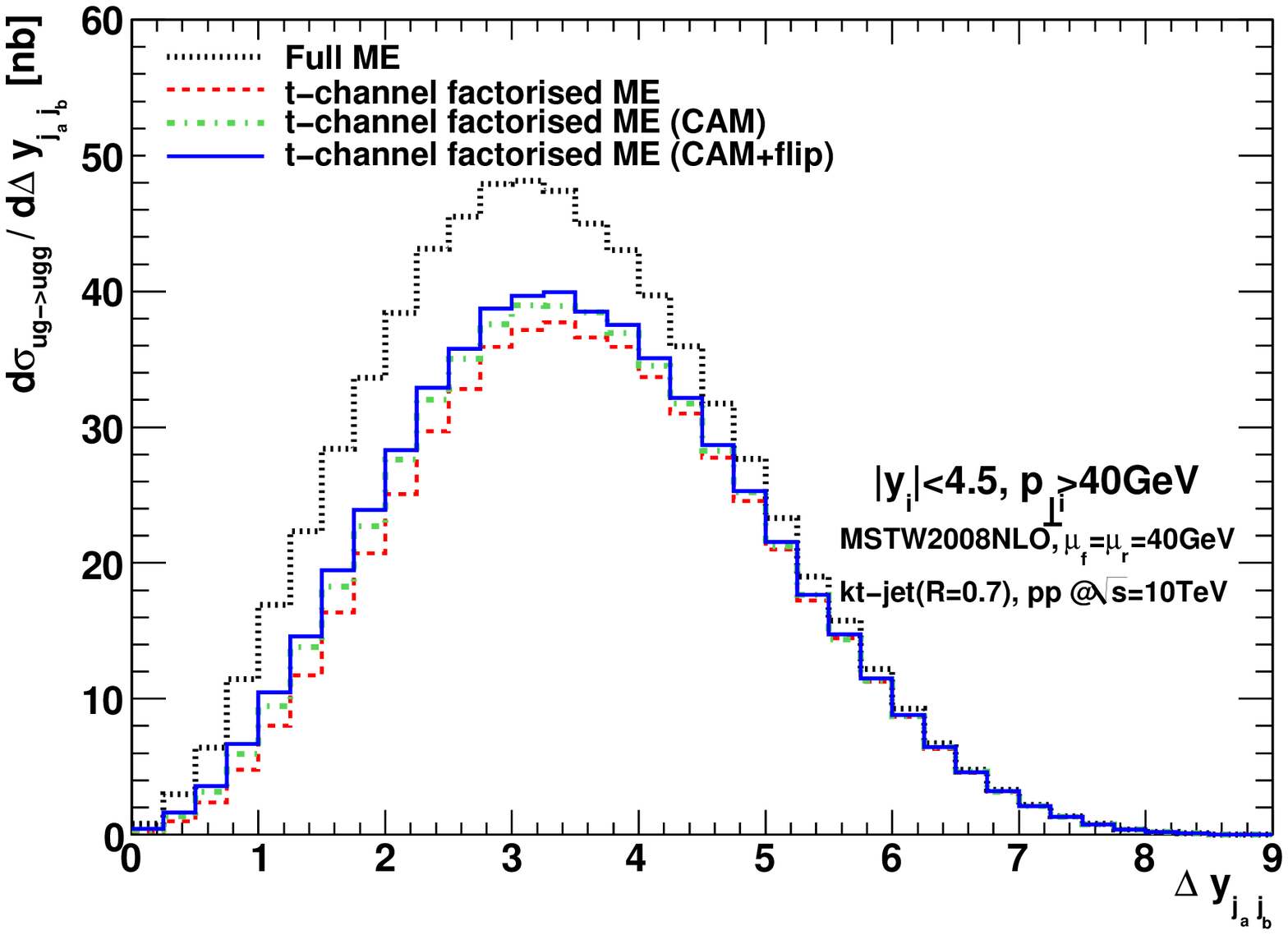}

  (a) \hspace{6cm} (b)
  \caption{(a) The angular distribution between and (b) the distribution of
    the rapidity difference of the forward and backward jet for $qg\to qgg$
    at the LHC.  The coloured lines display the incremental approximations
    discussed in Ref.\cite{Andersen:2009he}, with the blue line taking into
    account both the \emph{CAM} and the $t$-channel pole from amplitudes with
    a flip in the gluon helicity.  The solid black line is the full
    tree-level result taken from Madgraph~\cite{Alwall:2007st}.}
  \label{fig:qg3j}
\end{figure}

With the universality of the description of the $t$-channel pole established
as that of a current-current scattering, we can start approximating further
gluon emission. This is achieved by extracting a gauge-invariant effective
vertex, which takes into account soft emission from the basic $2\to2$
scattering process plus an emission from a connecting $t$-channel gluon. This
form guarantees the right MRK limit of all the amplitudes with $n$
partons. As an example, we compare in Fig.~\ref{fig:qg3j} the description of
the process $ug\to ugg$ at the LHC at the lowest order of the approximation
and the full tree-level matrix element. 
Obviously, the point is not just to reproduce results, where these can
already be calculated through other means. But by illustrating how well the
approximations work, where the results can be obtained through standard
approaches, we hope to instill trust in the many predictions of the
formalism, where these currently cannot be obtained reliably through other
approaches. The quality of the approximation of hard emissions is much better
than any other approach allowing for all-order estimates to be
obtained. Furthermore, the all-order formulation allows for matching to full
fixed order results, where these can be calculated\cite{Andersen:2008ue}.

Virtual corrections are approximated according to the \emph{Lipatov Ansatz},
which has been proved to get correct even the sub-leading logarithm of the
virtual corrections for the $n$-parton scattering
amplitude\cite{Bogdan:2006af}. The sum over real and virtual corrections is
finite at each order in perturbation theory. And the expressions are
sufficiently simple that the all-order result can be calculated as an
explicit sum over the explicit integration of $n$-particle phase space. So
all momenta are kept exclusive, and any analysis can be performed on these.
Full details may be found in~\cite{Andersen:2009nu,Andersen:2009he}, where we
also discuss the description of $W,Z,h$+jets.

\section{Summary}
\label{sec:summary}

We have discussed the very basic observation of the exact factorisation of
the $t$-channel pole in some $2\to2$ QCD processes. This forms the foundation
of the scheme introduced in Ref.\cite{Andersen:2009nu,Andersen:2009he}, which
allows for the construction of an all-order sum of parton emissions from
simple $2\to2$-processes constructed as a sum over explicit phase space
integrations over $n$-particle phase space. The sum tends to the exact
all-order scattering amplitude in the limit of large invariant mass between
all particles. This is the limit where the emission leads to additional jet
formation. A reliable description of hard emission is not only desirable, but
often critical for LHC analyses. The all-order summation of hard parton
emissions allow for the direct study of e.g.~jet veto efficiencies in gluon
fusion $hjj$ (see Ref.\cite{Andersen:2008gc}), where the inclusion of
several, hard emissions is crucial in reaching a stable prediction. Recently,
the analytic understanding of the all-order amplitude was used in designing
cuts and observables\cite{Andersen:2010zx}, which will significantly aid in
the extraction of the $CP$-properties of any Higgs boson created through
gluon fusion.

\bibliographystyle{JHEP}
\bibliography{jetpapers}

\end{document}